\documentclass[a4paper,fleqn]{cas-dc}

\usepackage[authoryear,longnamesfirst]{natbib}

\usepackage{tikz}
\usepackage{float}
\usetikzlibrary{arrows}
\usepackage{units}
\usepackage{setspace}
\usepackage{tabularx}
\usepackage{booktabs}
\usepackage{subfig}
\usepackage{todonotes}
\usepackage{nicefrac}
\usepackage{color}

\def\tsc#1{\csdef{#1}{\textsc{\lowercase{#1}}\xspace}}
\tsc{WGM}
\tsc{QE}

\definecolor{orange}{RGB}{255,127,0}
\definecolor{greeen}{rgb}{0.12, 0.3, 0.17}
\definecolor{dgreen}{RGB}{34,139,34}
\definecolor{dblue}{RGB}{10,25,150}

\begin{document}
\let\WriteBookmarks\relax
\def\floatpagepagefraction{1}
\def\textpagefraction{.001}

\shorttitle{Energy absorption of sustainable lattice structures under impact loading}    

\shortauthors{S. Bieler, K. Weinberg}  

\title [mode = title]{Energy absorption of sustainable lattice structures under impact loading}  

\tnotemark[1] 

\tnotetext[1]{} 

%

\author[1]{Sören Bieler}[orcid=0000-0002-1358-8267]

\cormark[1]

\fnmark[]

\ead{soeren.bieler@uni-siegen.de}

\ead[url]{}

\credit{Writing – original draft, Conceptualization, Software, Investigation,  Visualization, Validation}

\affiliation[1]{organization={Department of Mechanical Engineering, Festkörpermechanik, Universität
Siegen},
            addressline={Paul-Bonatz-Straße 9-11}, 
            city={Siegen},
            postcode={57076}, 
            state={NRW},
            country={Germany}}

\author[1]{Kerstin Weinberg}[orcid=0000-0002-2213-8401]

\fnmark[]

\ead{kerstin.weinberg@uni-siegen.de}

\ead[url]{}

\credit{Writing – review and editing, Methodology, Formal analysis, Supervision}


\cortext[1]{Corresponding author}

\fntext[1]{}


\begin{abstract}
Lattice structures are increasingly used in various fields of application due to the steady growth of additive manufacturing technology. Depending on the type of lattice, these structures are more or less suitable for energy absorption due to the deformation of diagonal struts. The energy absorption properties depend significantly on the type of the selected lattice structure and its density, material properties, printing process, and post-treatment. Here, five lattice types (\textsc{Octet}, \textsc{BFCC}, \textsc{Diamond}, \textsc{Truncated Octahedron} and \textsc{Rhombicuboctahedron}) with different volume fractions are compared. Stereolithography is used to print the different lattices made from liquid resin. This allows good results to be achieved with tiny structures. In particular, the sustainability of energy-absorbing structures plays a significant role in many processes to withstand multiple loads. The lattice structures are made of TPU resin and offer different energy absorption properties without being destroyed under load. The structures are loaded abruptly using the Split-Hopkinson pressure bar test in a modified setup. From the measured strain pulses, we can calculate how much of the applied energy was absorbed by the different structures. 
\end{abstract}


\begin{highlights}
\item 
\item 
\item 
\end{highlights}



\maketitle



\section{Introduction}\label{sec:introduction}

Cellular materials can undergo large deformations under mechanical load and yet return to their original state when the load is released. This makes them suitable for non-destructive, reversible, and sustainable energy absorption. They are used in protectors that cushion sudden collisions, as packaging, or simply as a soft cushioning base.

Natural cellular materials are, for example, the skin of the pomelo, which survives falls from ten meters and absorbs up to 90\.\% of the energy on impact. Here, the cell structure is irregular, uneven, and challenging to reproduce in engineering applications.
On the other hand, there are foams whose cells or pores are irregularly arranged but have almost uniform macroscopic properties. Because they are cheap to produce and behave (visco)elastically, polymer foams are mainly used in protective equipment such as helmets or knee pads.

With the advent of additive manufacturing, it has become possible to create structures of almost any complexity. This raises the question of whether much more effective cellular structures than foam can be constructed, for example, those inspired by the crystal lattice properties of solid metals. Numerous lattice structures have now been designed and also mechanically tested; for a status of research and a classification of the lattices into subcategories, please refer to the review \cite{yin2023review}. However, most experimental investigations were carried out under static load \cite{song2019octet,nakarmi2024role}, although dynamic load conditions are undoubtedly more important for practical use as energy dissipators. The authors \cite{ling2019mechanical} realized dynamic compression tests of polymeric \textsc{Octet} by a drop mass of 6.5 kg and an impact speed of 3 m/s.

We use a split Hopkinson bar (SHB) to load five truss-like lattice structures by impact, measure their deformation, and derive the stress-strain curves. The lattice structures are made of thermoplastic polyurethane (TPU) and behave viscoelastically. They return to their original state after a specific recovery time and can withstand multiple loads. The SHB apparatus is modified compared to the one used in \cite{kang2023green}, as only the first impulse in the incident bar is considered for our approach. From the measurement results, we derive statements on the different energy dissipation due to the lattice shape. To ensure comparability, the lattice structures each have the same mass.

Our paper is organized as follows: Next, in Section~\ref{sec:lattice_structure_design},  we introduce the investigated lattice structures, discuss their geometry, material, and the additive manufacturing process. The conducted experiments will be discussed in Section~\ref{sec:experiments}, where we also 
present our results, i.e. the strain-over-time diagrams, the frequency spectra and the specific energy absorption for each lattice structure. In the following Section~\ref{sec:simulation} we present and discuss numerical simulations of the experiments.
We conclude with a discussion of our findings in Section~\ref{sec:discussion_summary}.

%
\section{Lattice structures} \label{sec:lattice_structure_design}

In this section, we explain the choice of lattices investigated and describe the preparation of the specimens.

\subsection{Classification}

\begin{figure*}[htb]
\centering
\resizebox{1\textwidth}{!}{
\begin{tikzpicture}
\draw (0,0) node[anchor=south west]{
\includegraphics[width=1\textwidth]{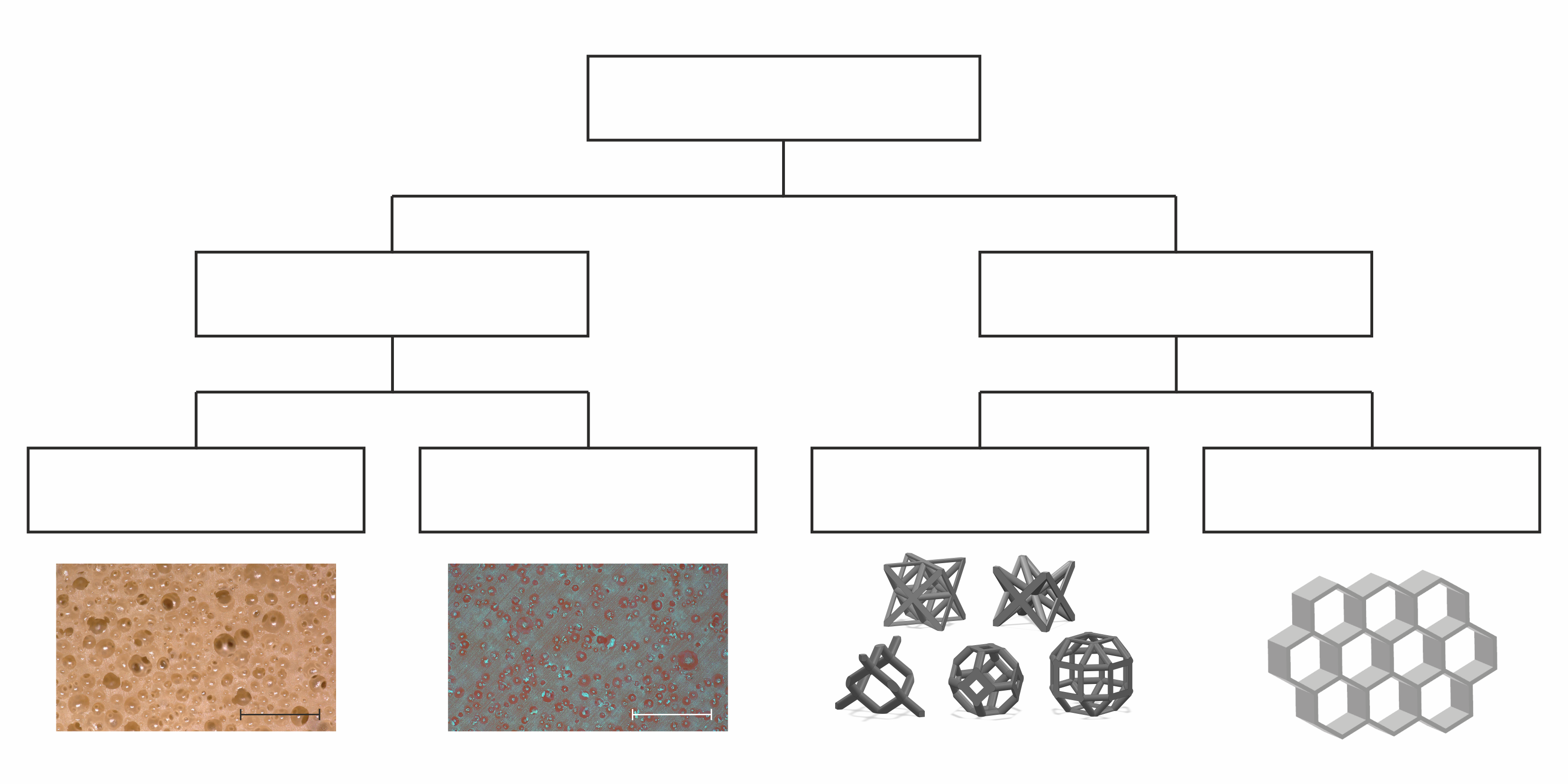}};
\draw(7.1,7.45) node[anchor=south west]{\small{cellular (micro) structure}};
\draw(11.9,5.25) node[anchor=south west]{\small{regular structure}};
\draw(3,5.25) node[anchor=south west]{\small{irregular structure}};
\draw(1.5,3.1) node[anchor=south west]{\small{open-cell}};
\draw(5.7,3.1) node[anchor=south west]{\small{closed-cell}};
\draw(10.1,3.1) node[anchor=south west]{\small{open-cell}};
\draw(14.4,3.1) node[anchor=south west]{\small{closed-cell}};
\draw(7.1,0.8) node[anchor=south west]{\textcolor{white}{\tiny{1000 $\mu$m}}};
\draw(2.75,0.8) node[anchor=south west]{\tiny{1000 $\mu$m}};
\end{tikzpicture}}
\caption{Classification of cellular structures  \citep{park2022design}}
\label{fig:classification}
\end{figure*}
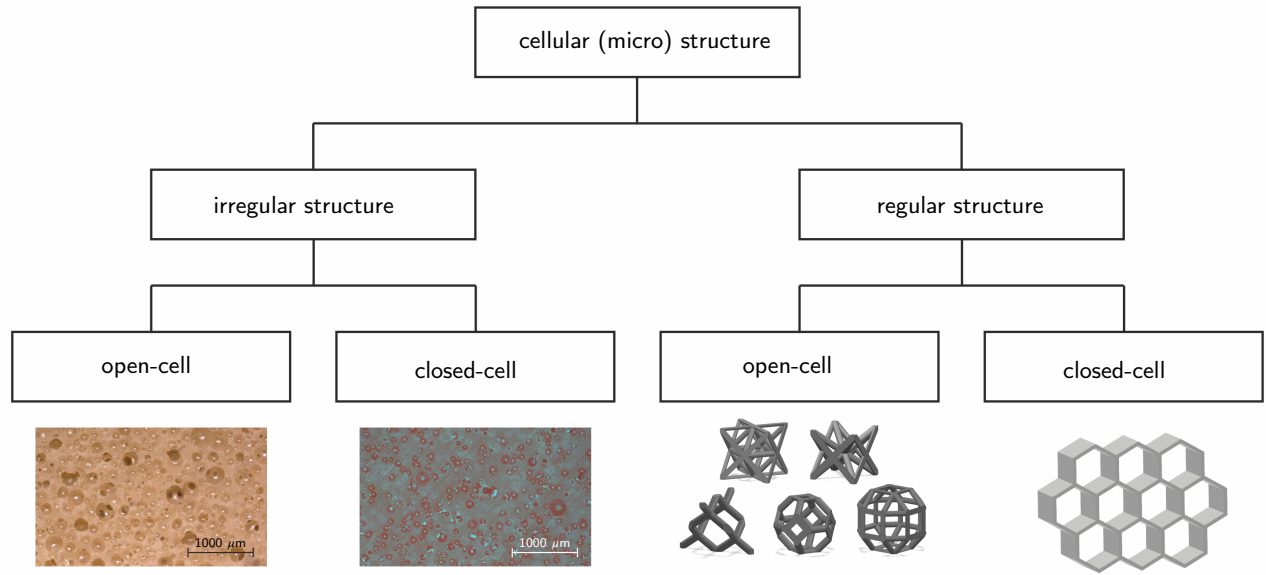

Cellular materials are initially divided into irregular structures and regularly repeating arrangements. Both categories can be subdivided because cellular structures can be open-celled or closed-celled; see Fig.~\ref{fig:classification}.

Irregular microstructures have cell or pore sizes that are gradually or stochastically distributed, such as in bark, cork, or foam. Closed-cell foams usually have spherical pores, while in open-cell foams, the pores grow into each other and therefore, the (inverse) ligament structure is better described, cf. \cite{ashby1997cellular,bogunia2016experimental}. In natural open-cell foams, four struts usually meet in one node, cf. \cite{gong2005compressive}.


For regular structures, closed-cell formations are mainly obtained by continuing the two-dimensional structure in the third direction. This is how honeycomb cores for laminate structures are created, for example. Here, work on optimizing the cell structure is relatively rare, and if so, then in connection with auxetic or hierarchical lattices, cf. \cite{walkowiak2022numerical}. Three-dimensional regular microstructures are composed of lattice elements in the form of beams or shells. These lattice elements can meet in any number at the nodes. This connectivity, i.e., the different number of struts per cell and node, significantly affects the overall mechanical behavior.

There are many open-cell lattice structures, each with advantages and disadvantages. As the use of 3D printers advances, it becomes increasingly easier to produce such structures and study their properties. Of a large number of possible structures, only a selection of those that were last examined under dynamic conditions will be mentioned here: 
the body-centered cube (\textsc{BCC}) based on cubic crystal system \cite{tancogne2018stiffness}, \textsc{Octet},consisting of tetra and octahedron \cite{tancogne2016additively}, \textsc{re-entrant cube} and \textsc{Diamond} \cite{ozdemir2016energy} and \textsc{FCCZ} \cite{bieler2021investigation}.

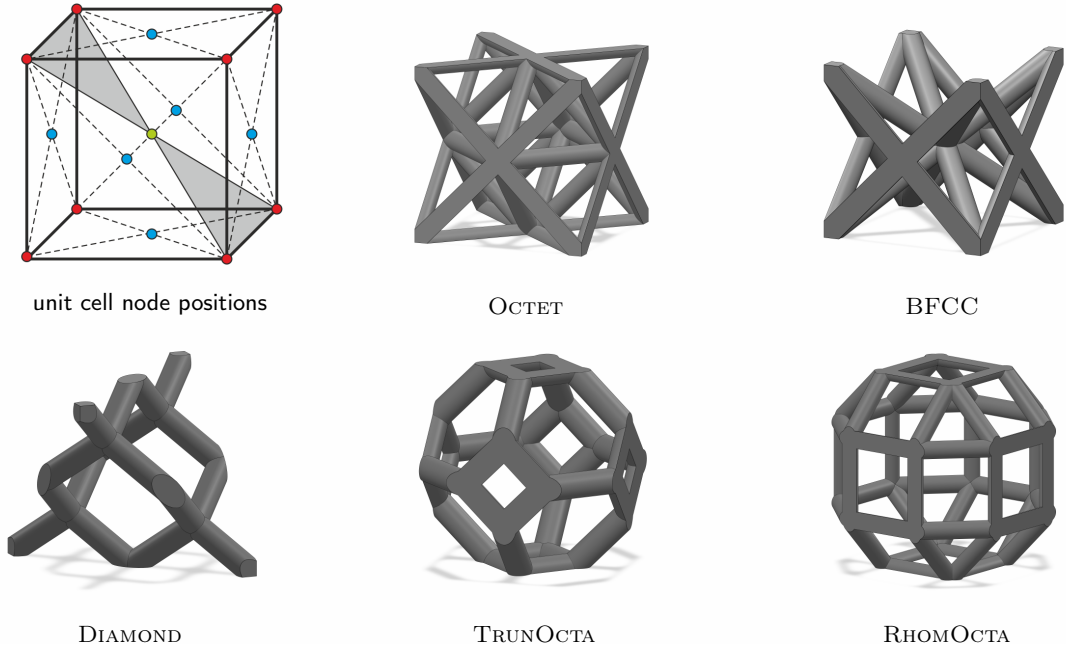
\begin{figure*}[htb]
\centering
\resizebox{1\textwidth}{!}{
\begin{tikzpicture}
\draw (0,0) node[anchor=south west]{
\includegraphics[width=1\textwidth]{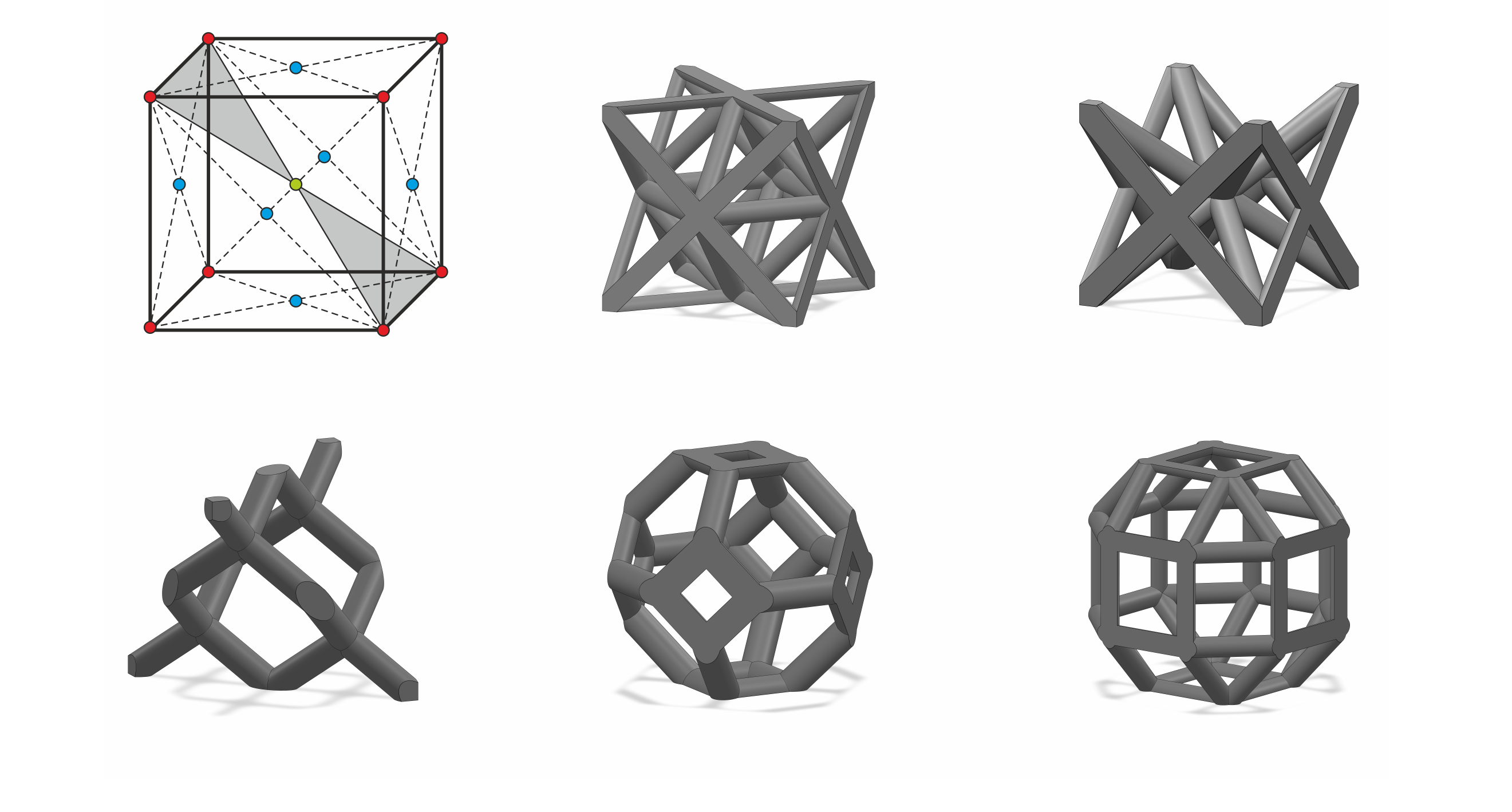}};
\draw(1.8,4.8) node[anchor=south west]{unit cell node positions};
\draw(7.9,4.8) node[anchor=south west]{\textsc{Octet}};
\draw(13.5,4.8) node[anchor=south west]{\textsc{BFCC}};
\draw(2.4,0.4) node[anchor=south west]{\textsc{Diamond}};
\draw(7.7,0.4) node[anchor=south west]{\textsc{TrunOcta}};
\draw(13.2,0.4) node[anchor=south west]{\textsc{RhomOcta}};
\end{tikzpicture}}
\caption{Unit cells of the investigated lattice types}
\label{fig:unit_cells}
\end{figure*}

Here we compare different beam-based lattice structures. This limitation makes it possible to produce different truss-lattice structures with the same mass by adjusting the radius of the struts or  beams accordingly. The five types of lattice we investigated are:
\begin{itemize}
  \item the conventional octet truss lattice composed of tetrahedra and octahedra: \textsc{Octet}
  \item the face-centered cubic lattice with an additional body-centered node: \textsc{BFCC}
  \item the diamond lattice squeezed to fit in a cubic cell: \textsc{Diamond}
  \item the  convex polyhedra truss arising from a octahedron by cutting of the vertices (truncated octahedron): \textsc{TrunOcta}
  \item the convex small rhombicuboctahedron truss: \textsc{Rhom\-Octa} 
\end{itemize} 
The corresponding unit cells are shown in Fig. \ref{fig:unit_cells}; details of the lattices' connectivitiy are summarized in Table~\ref{tab:struts_nodes}. Since all generated unit cells are symmetric, nodes can be shared at the corners, along the edges and on the faces of the unit cell. 

The \textsc{Octet} is an established and well-investigated lattice that may serve as a reference to a certain extent. The \textsc{BFCC} structure is a combination of the common crystal-like lattices \textsc{BCC} and \textsc{FCC}. By merging the two basic structures, a more complex structure is obtained, from which we would expect a relatively stiff answer. Conversely, the \textsc{TrunOcta} and the \textsc{RhomOcta}
are both lattices with a `hole' in the center. Here one may expect large deformations but also failure because of buckling. 
The \textsc{Diamond} lattice is specifically designed for being stiff. Here are basically four cheval-de-frise-like trusses combined to achieve a deformation-stopping structure. This gives a simple, loosely packed lattices with --~at equal mass~-- relatively thick struts.


In an attempt to quantify these characteristics, the lattice's topological properties are described by the Maxwell number $M$,\cite{maxwell1864calculation}. The Maxwell number $M$ classifies the dominant loading regime. It is calculated from the $n$ nodes and the $s$ struts connected to the nodes. The lattice's topological properties and dominant loading regime can partially be described by the Maxwell number $M$. It is calculated from the $n$ nodes and the $s$ struts connected to the nodes, cf. \cite{deshpande2001effective}, 
\begin{align} \label{eq:maxwell}
    M = s - 3\,n + 6. 
\end{align}
For $M \geq 0$, the structure has a stretch-dominated behavior. A bending-dominated behavior corresponds to structures with $M< 0$. Table \ref{tab:struts_nodes} provides the data for the lattices investigated here, with the result, that only the popular \textsc{Octet} structure ($M = 0$) has a stretch-dominated behavior. All other structures have a bending-dominated behavior: {BFCC} ($M = -9$), \textsc{Diamond} (M=-2), 
\textsc{RhomOcta} ($M = -18$) and the \textsc{TrunOcta} ($M = -30$).

\begin{table*}[htb]
\caption{Number of struts and nodes at different positions of the lattice unit cells}
\label{tab:struts_nodes} 
\begin{tabularx}{\textwidth}{lXXXXr} \toprule
\noalign{\smallskip}
 & struts &  & nodes & \\
 \noalign{\smallskip}
 &  & corner & center & face & $M$\\
\noalign{\smallskip}
\hline
\noalign{\smallskip}
\textsc{Octet} & 36 & 8 & - & 6 & 0\\
\noalign{\smallskip}
\textsc{BFCC} & 24 & 8 & 1 & 4 & -9\\
\noalign{\smallskip}
\textsc{Diamond} & 16 & 4 & - & 4 & -2\\
\noalign{\smallskip}
\textsc{RhomOcta} & 48 & - & - & 24 & -18\\
\noalign{\smallskip}
\textsc{TrunOcta} & 36 & - & - & 24 & -30\\
\noalign{\smallskip}
\hline
\end{tabularx}
\end{table*}

We remark that the Maxwell number is a simple characteristic derived for classical trusses.  Large deformations of the lattice which include buckling and folding of layers are not weighted. We  also remark that for an irregular, stochastic microstructure the material characterization requires sufficiently large samples and/or a large number of experiments. This is in contrast to ordered structures, which consist of periodically repeating unit cells. Here, the material response can also be determined on small samples that fluctuate solely due to manufacturing inaccuracy.

\subsection{Specimen design} \label{subsec_specimen_design}

The key parameter for cellular materials is the volume fraction $f_\text{V}$. For periodically continued lattices, it can be calculated from the unit cell, and it represents the ratio between the volume of a cube $V_{\text{ref}}$ with the edge length $L$ and the actual material volume of the lattice $V_\text{lattice}$. The latter can be estimated from the number of struts and their length, but the exact value must be obtained from CAD drawings or by weighting the specimen.
\begin{align}
f_\text{V} = \frac{V_\text{lattice}}{V_{\text{ref}}} \qquad \text{with} \qquad V_{\text{ref}}=L^3\label{eq:volume_fraction}
\end{align}
The edge length  of a unit cell is set to $L$ = 4.5 mm for all lattices.
Three different volume fractions $f_\text{V}$ are examined, namely:
\begin{itemize}
  \item $f_\text{V}$ = 0.2\,,
  \item $f_\text{V}$ = 0.3\,,
  \item $f_\text{V}$ = 0.4\,.
\end{itemize}

\begin{figure*}[htb]
\begin{tikzpicture}
\draw (0,0) node[anchor=south west]{
\includegraphics[width=1.0\textwidth]{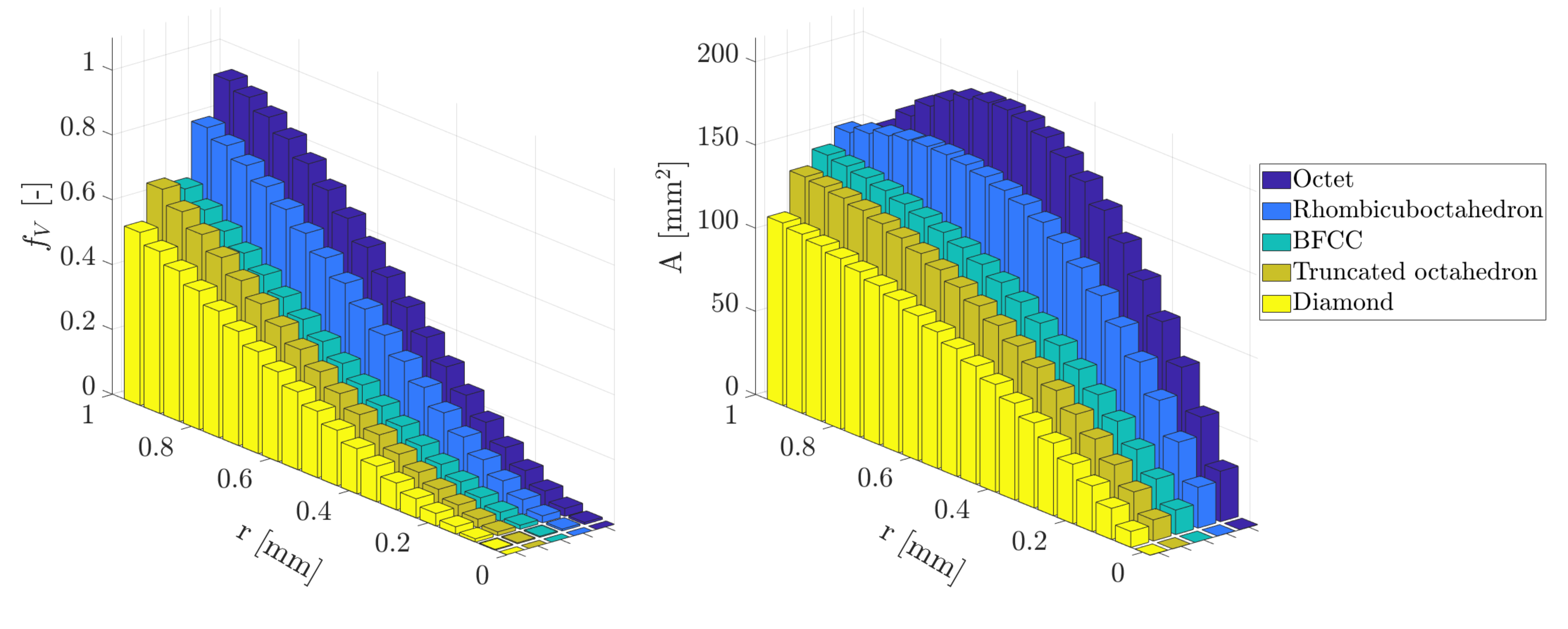}};
\draw(0.5,0) node[anchor=south west]{a)};
\draw(7.8,0) node[anchor=south west]{b)};
\draw(5,-0.4) node[anchor=south west]{ }; 
\end{tikzpicture}
\caption{Volume fraction $f_{\text{V}}$ a) and surface area b) for different strut radii and a constant edge length of $L=4.5$ mm per unit cell}
\label{fig:volume_fraction}
\end{figure*}
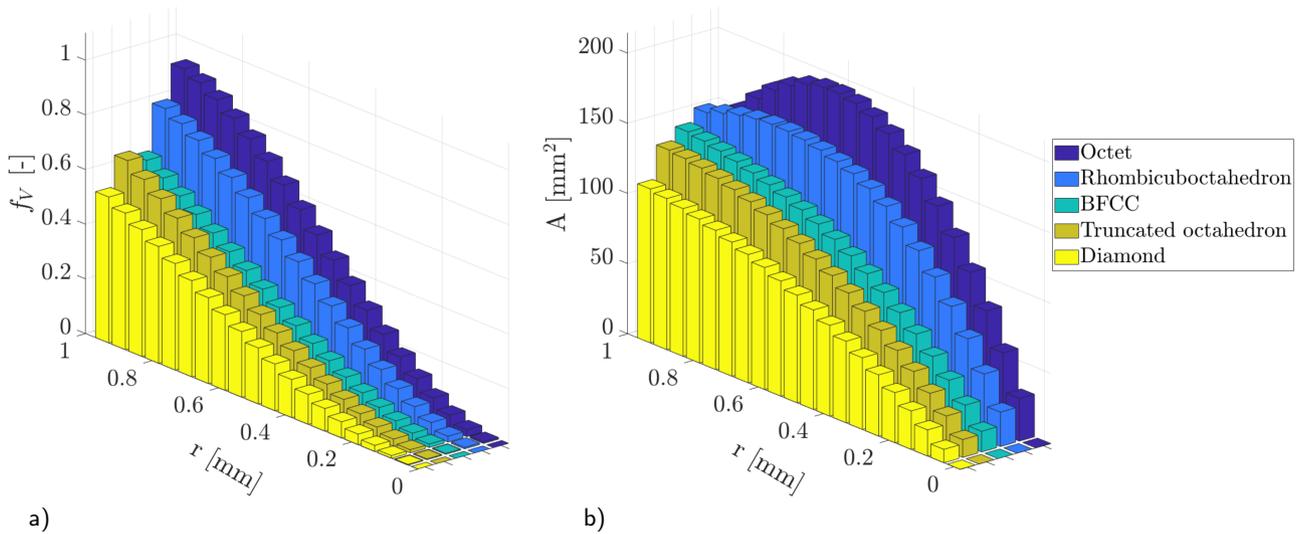

We remark that it is not trivial to prescribe a value of $f_\text{V}$ for the different types of lattices. Fig. \ref{fig:volume_fraction} shows the volume fraction as well as the surface area  as a function of the  struts' radius.
It can be seen that the radius for the \textsc{Diamond} and \textsc{TrunOcta} structure must be higher than that of the \textsc{Octet} in order to generate the same $f_\text{V}$. Raising the volume fraction further would result in structures which are not truss-like anymore.

To design the specimen,  $3\times 3\times 3$ unit cells are combined to a lattice structure, see Fig. \ref{unit_cell_final_specimen}~b).
Table \ref{tab:volume_fraction} in the Appendix summarizes the strut radii of the 15 different structures for the prescribed  volume fractions.

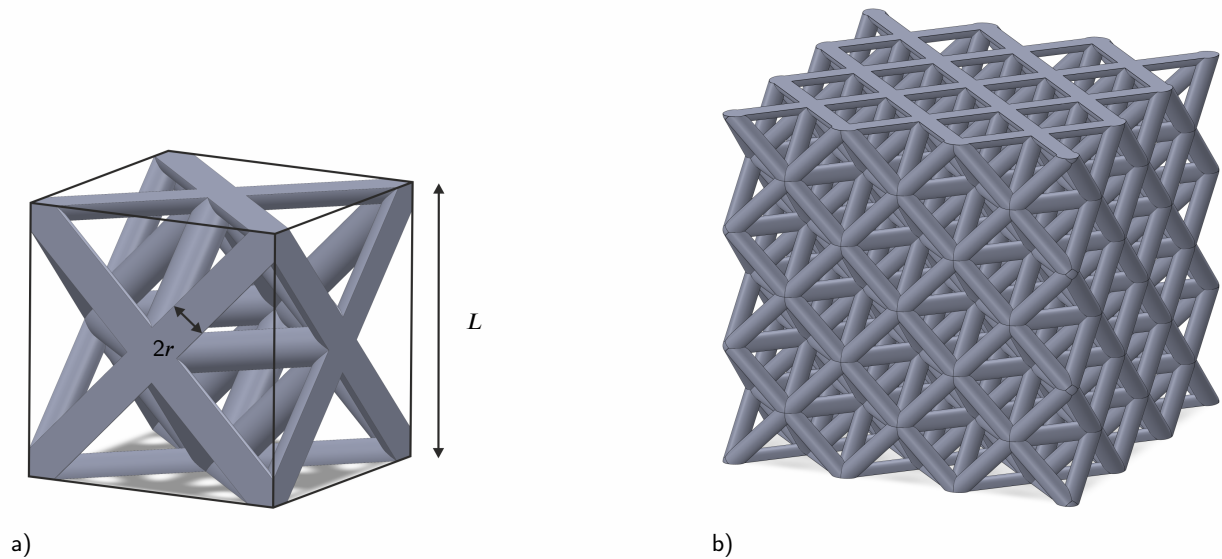
\begin{figure*}[htb]
\centering
\resizebox{1\textwidth}{!}{
\begin{tikzpicture}
\draw (0,0) node[anchor=south west]{
\includegraphics[width=1\textwidth]{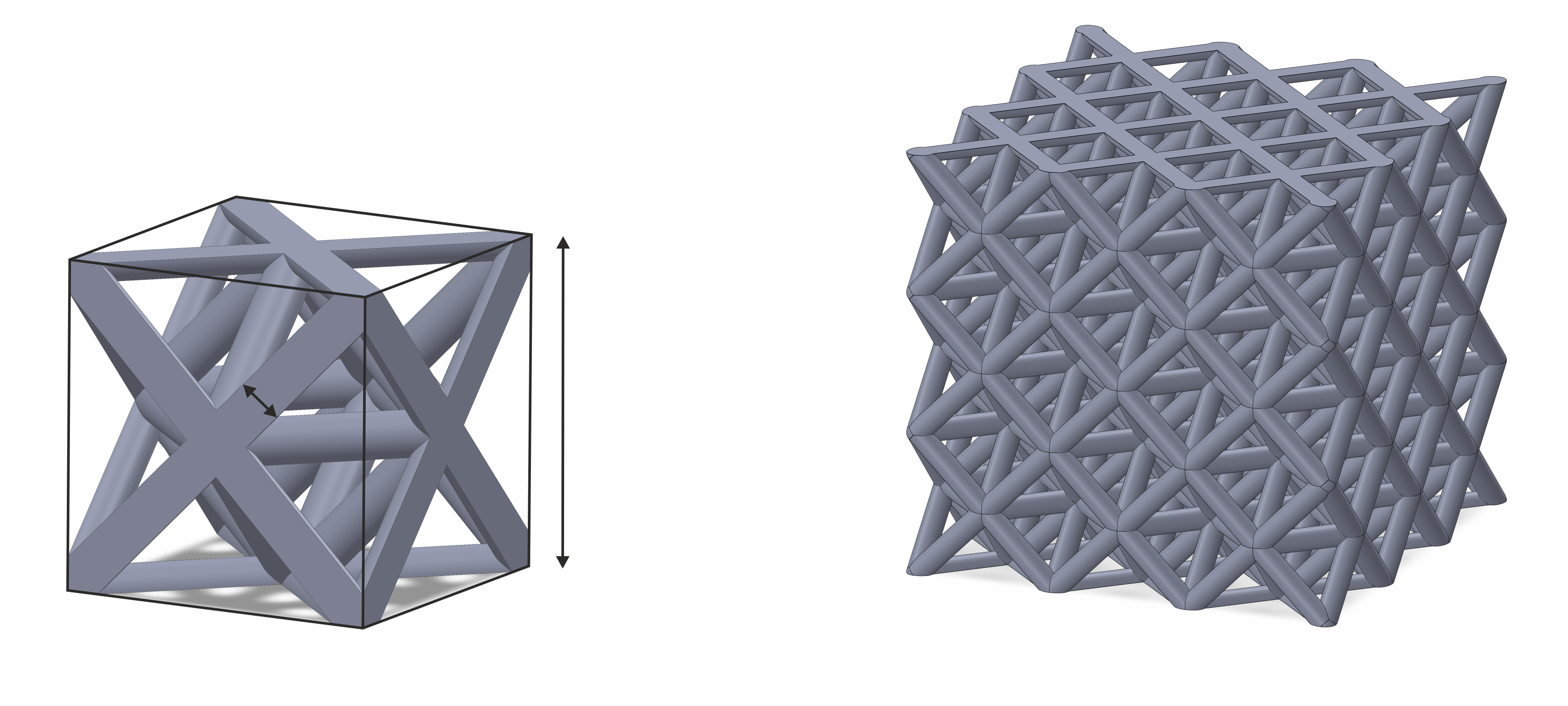}};
\draw(0.5,0.2) node[anchor=south west]{a)};
\draw(9.9,0.2) node[anchor=south west]{b)};
\draw(2.4,2.9) node[anchor=south west]{2$r$};
\draw(6.6,3.25) node[anchor=south west]{$L$};
\end{tikzpicture}}
\caption{\textsc{Octet} structure: a) unit cell and b) final specimen (3x3x3 unit cells)}
\label{unit_cell_final_specimen}
\end{figure*}

\subsection{Printing process} \label{subsec_printing_process}
The specimens are produced by additive manufacturing using a \textsc{Digital Light Processing} printer, specifically a stereolithography (SLA) process. In this process, the liquid synthetic resin is cured layer by layer by a light source. The display's high resolution in the printer achieves a voxel size of 35\,$\mu$m due to its relatively small installation space. This makes it possible to create very fine lattice structures.

However, in the case of complex structures such as lattices, care must be taken to ensure that the support structures are placed outside the actual structure, as their subsequent removal is difficult, cf. \cite{bieler2023behavior}. Due to the small dimensions of the tested specimens, the structures are stable enough to be printed without any support structures. As no mechanical post-processing is required, the structures have very uniform and clean surfaces. After the printing process, the specimens are freed from liquid resin residues using an isopropanol bath. The washing process is followed by curing the specimens under material-specific parameters (temperature and UV light).

\subsection{Material}
The lattice structures should withstand multiple impacts without being destroyed. That requires to use a material that has a reversible behavior and does not deform plastically. 
In general, a wide variety of SLA resins are available on the market. They range from hard materials like acrylate-based resins like the to more ductile impact-resistant resins like polypropylene-based (PP) and to rubbery materials like thermoplastic polyurethanes (TPU). Brittle or tough materials are hardly suitable for impact applications because they tend to break under the sudden load. In order to produce sustainable lattice structures, we have selected various rather soft materials.

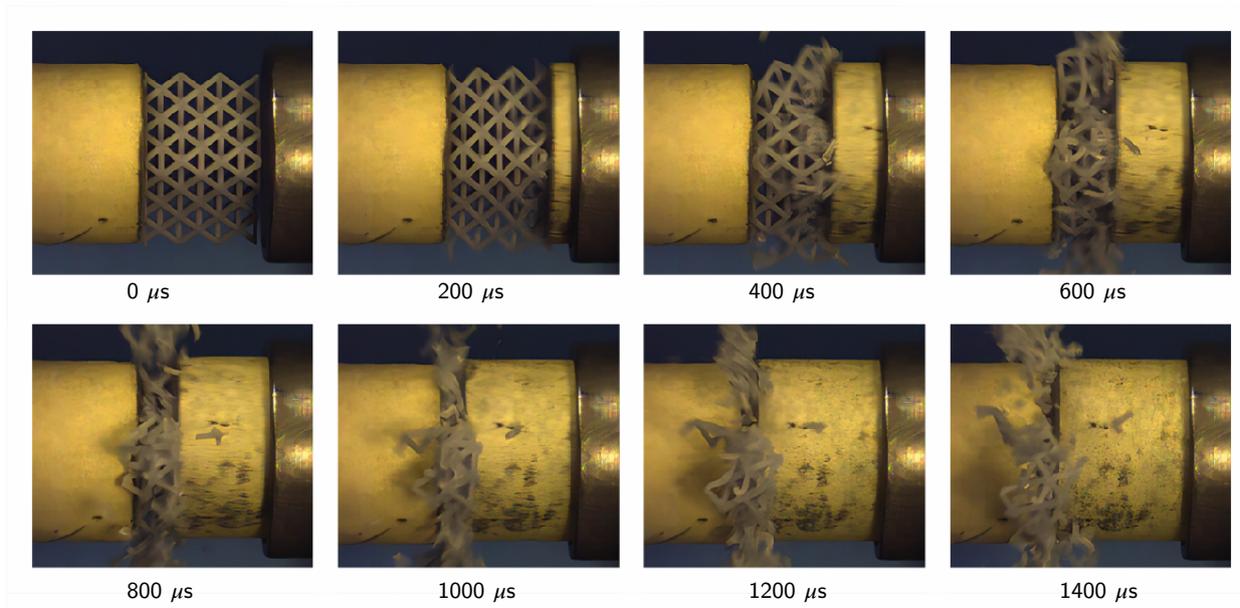
\begin{figure*}[htb]
\centering
\resizebox{0.97\textwidth}{!}{
\begin{tikzpicture}
\draw (0,0) node[anchor=south west]{
\includegraphics[width=1\textwidth]{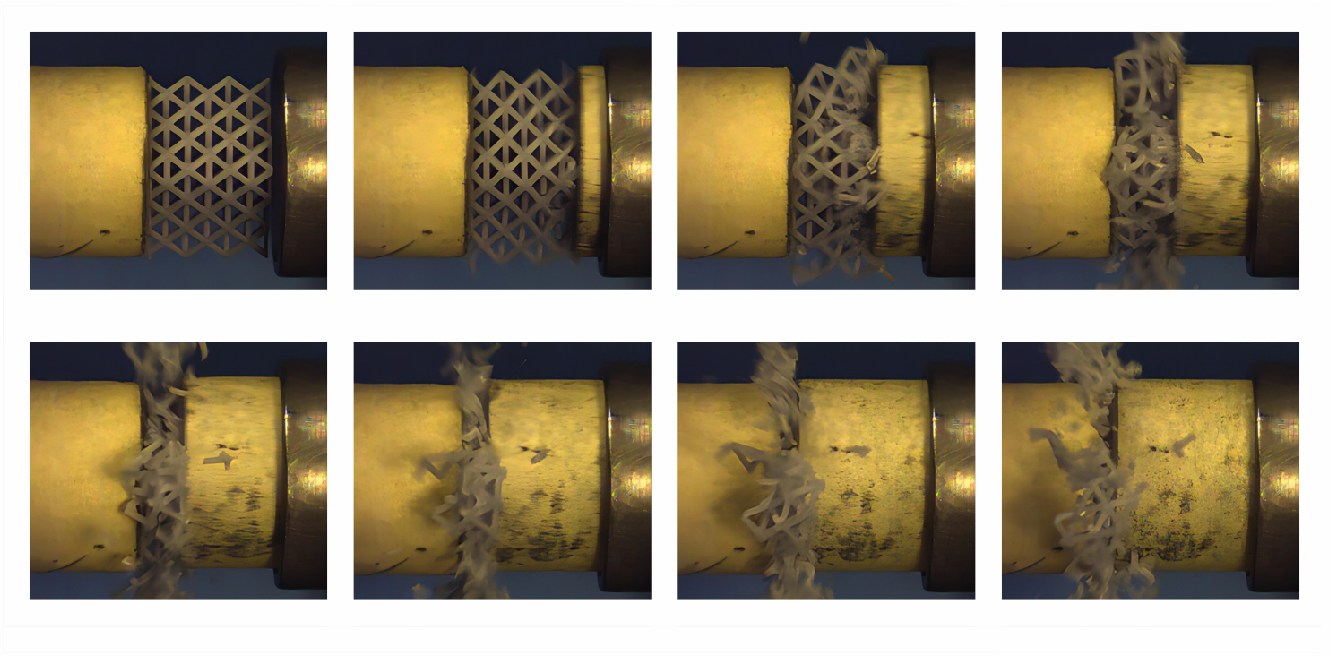}};
\draw(1.7,4.4) node[anchor=south west]{0 $\mu$s};
\draw(6,4.4) node[anchor=south west]{200 $\mu$s};
\draw(10.3,4.4) node[anchor=south west]{400 $\mu$s};
\draw(14.6,4.4) node[anchor=south west]{600 $\mu$s};
\draw(1.7,0.25) node[anchor=south west]{800 $\mu$s};
\draw(6,0.25) node[anchor=south west]{1000 $\mu$s};
\draw(10.3,0.25) node[anchor=south west]{1200 $\mu$s};
\draw(14.6,0.25) node[anchor=south west]{1400 $\mu$s};
\end{tikzpicture}}
\caption{Different states of deformation during the experiment with specimen  made of ABS like material (here \textsc{Octet} $f_\text{V}$ = 0.2)}
\label{specimen_break_ABS}
\end{figure*}

In \cite{bieler2024Meccanicasubmitted} we investigated the same lattices structures under static compression. The specimens used there were made of acylic-based polymer resin (\textsc{Anycubic Tough Resin 2.0}) and reversibly compressed of more than 50\,\%. Doing the same here, under dynamic impact, leads to failure of the lattice structures, see Fig.~\ref{specimen_break_ABS}.
The structures still dampened the impact, but because of the desired sustainability, a destructive behavior does not match our goals.

Suitable materials for the application of multiple loads are those with a pronounced viscoelastic behavior. After the specimens have been relieved, the structure should return to its original configuration and can be loaded again. For the experiments carried out in this paper, the lattice structures were made of the TPU \textit{SuperFlex} of the company \textit{3DMaterials}. According to the manufacturer it has secant modulus of 32\,MPa evaluated according to \textsc{ASTM D 638}. The combination of reversibility and stiffness makes it possible to load the material repeatedly without any plastic deformation.

\section{Experiments} \label{sec:experiments}


The impact experiments were carried out on a SHB setup of PMMA bars. A conventional SHB consists of two long bars aligned along their axis. Both bars are made of the same material and have a circular cross-section; the specimen is clamped between the two bars.  A striker is accelerated by a gas gun to hit the first  (incident) bar, which induces a pressure pulse. This pressure wave travels through the incident bar and at the specimen it is in part reflected and in part transmitted to the second (transmission) bar.  The wave signal is recorded by strain gauges mounted in the center of both bars.

\subsection{SHB setup}
The dimensions of the SHB components can be found in the Appendix in Table \ref{tab:SHPB_dimensions}. The applicated strain gauges are wheatstone bridges in a full-bridge configuration (\textsc{HBM 3/350 XY31}) with a resistance of 350.0 $\omega \pm$  0.30 \% and a gauge factor of 2.0 $\pm$ 1.0 \%. In all performed experiments, the temperature of the laboratory was constant at 20$^\circ$ C. The applicated strain gauges have a length of approx. 3 mm, which allows a precise measurement of the propagating wave. With the \textsc{HBM GEN7t} data acquisition system, equipped with two bridge cards of type \textsc{GN411}, we are able to record data at a sample rate of 10$^{-6}$\,s$^{-1}$. Four channels per card thus allow us to record signals simultaneously at a total of eight measuring locations, each with a data acquisition rate of 1\,MS/s.

\subsection{Modified SHB experiment setup}
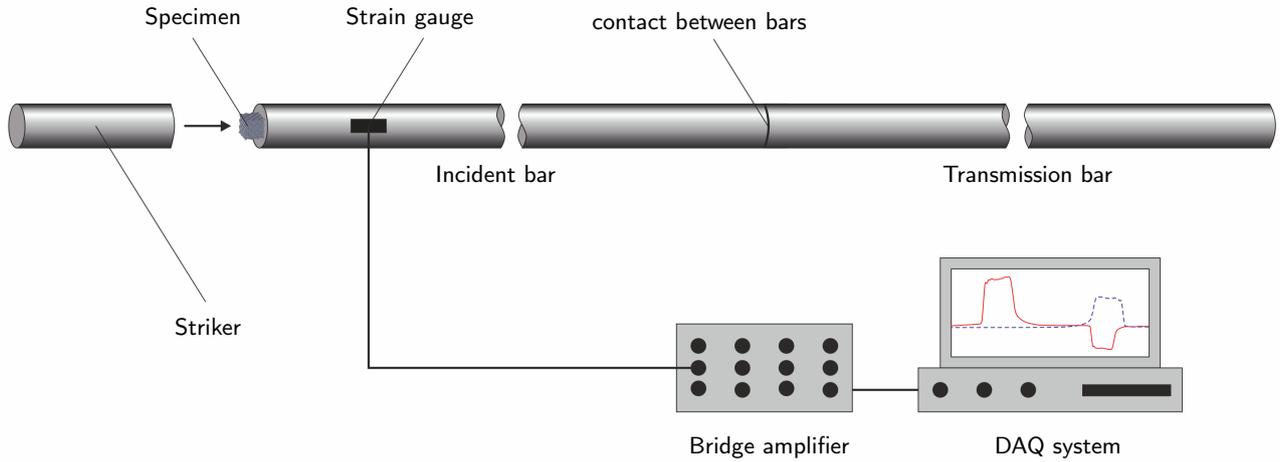
\begin{figure*}[htb]
\centering
\resizebox{1\textwidth}{!}{
\begin{tikzpicture}
\draw (0,0) node[anchor=south west]{
\includegraphics[width=1\textwidth]{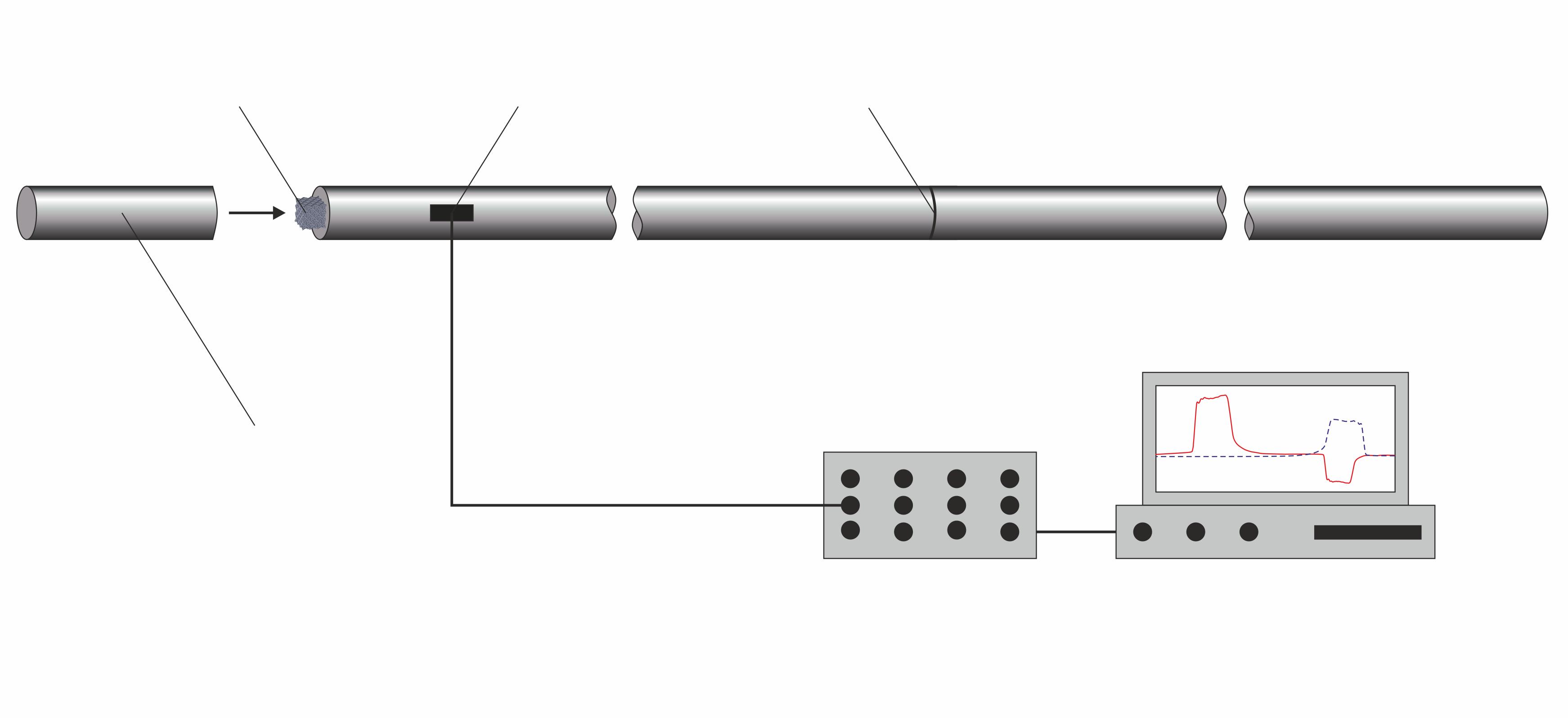}};
\draw(2.4,2.8) node[anchor=south west]{Striker};
\draw(4.7,6.9) node[anchor=south west]{Strain gauge};
\draw(9.3,1.15) node[anchor=south west]{Bridge amplifier};
\draw(8,6.9) node[anchor=south west]{contact between bars};
\draw(13.4,1.15) node[anchor=south west]{DAQ system};
\draw(2,6.9) node[anchor=south west]{Specimen};
\draw(5.9,4.85) node[anchor=south west]{Incident bar};
\draw(12.7,4.85) node[anchor=south west]{Transmission bar};
\end{tikzpicture}}
\caption{Modified SHPB Setup}
\label{fig:SHPB_modified}
\end{figure*}

After preliminary tests we decided to use the SHB setup in a modified way, see Fig. \ref{fig:SHPB_modified}. Now, the specimen is applied to the front face of the incident bar and the striker hits the specimen directly. The strain wave is measured at the gauge of the incident bar. 

Since only the induced pulses in the incident bar and the comparison when applying different lattice structures are essential for this paper, a signal prediction or correction, as in the conventional SHP experiment with a PMMA bar setup \cite{aghayan2022determination, bieler2021correction}, does not play a role. In this setup, the transmission bar ensures that the pulse is only transmitted and not reflected at the end of the incident bar. This prevents incoming and reflected pulses from mixing. The impact velocity of the striker is 15\,m/s. The experiments were recorded with a high-speed camera with a frame rate of 30,000 fps.

The investigation aims to determine to what extent the different lattice types can delay the load from the projectile over time while deforming reversibly. For reference, a series of tests is carried out without any specimen. Using the strain gauges, as described above, we determine the strain that is transferred into the system. By comparing the reference measurement and the actual experiments with applied specimen, we calculate the difference and the amount of energy absorbed.

\begin{figure*}[htb]
\centering
\resizebox{0.97\textwidth}{!}{
\begin{tikzpicture}
\draw (0,0) node[anchor=south west]{
\includegraphics[width=1\textwidth]{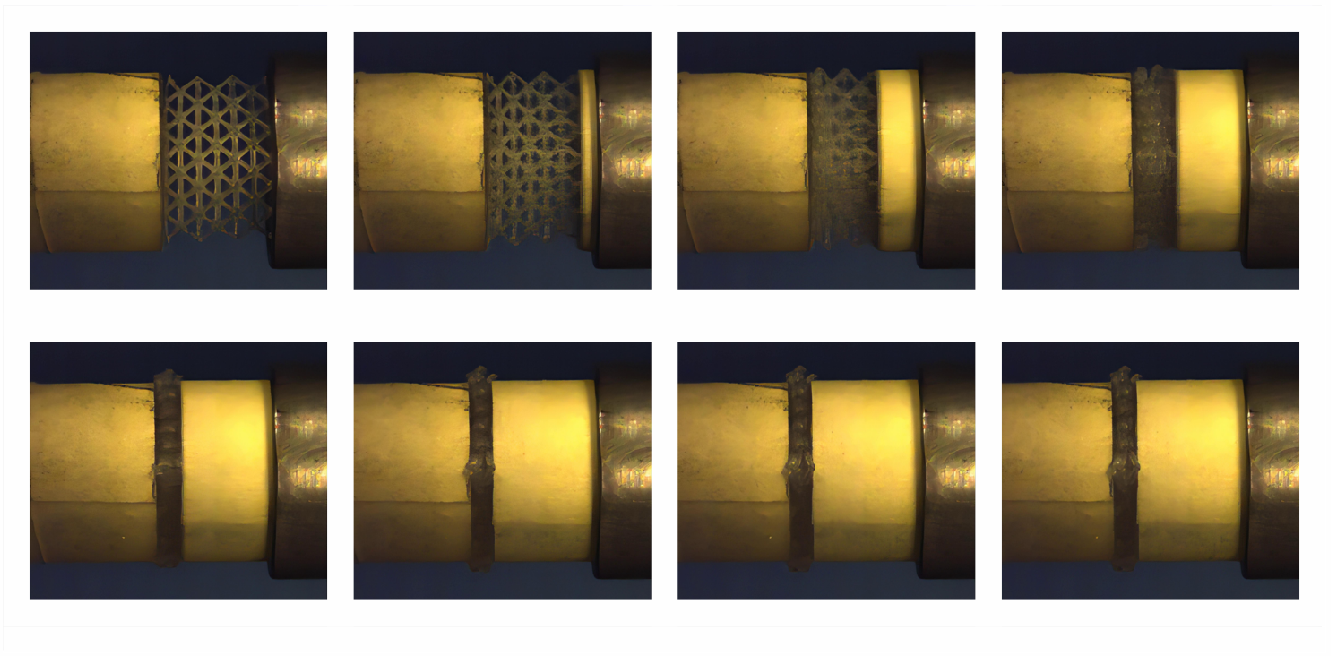}};
\draw(1.7,4.4) node[anchor=south west]{0 $\mu$s};
\draw(6,4.4) node[anchor=south west]{200 $\mu$s};
\draw(10.3,4.4) node[anchor=south west]{400 $\mu$s};
\draw(14.6,4.4) node[anchor=south west]{600 $\mu$s};
\draw(1.7,0.25) node[anchor=south west]{800 $\mu$s};
\draw(6,0.25) node[anchor=south west]{1000 $\mu$s};
\draw(10.3,0.25) node[anchor=south west]{1200 $\mu$s};
\draw(14.6,0.25) node[anchor=south west]{1400 $\mu$s};
\end{tikzpicture}}
\caption{Different states of deformation during the experiment with specimen  made of TPU like material (here \textsc{Octet} $f_\text{V}$ = 0.2)}
\label{specimen_break_TPU_octet}
\end{figure*}

\begin{figure*}[htb]
\centering
\resizebox{0.97\textwidth}{!}{
\begin{tikzpicture}
\draw (0,0) node[anchor=south west]{
\includegraphics[width=1\textwidth]{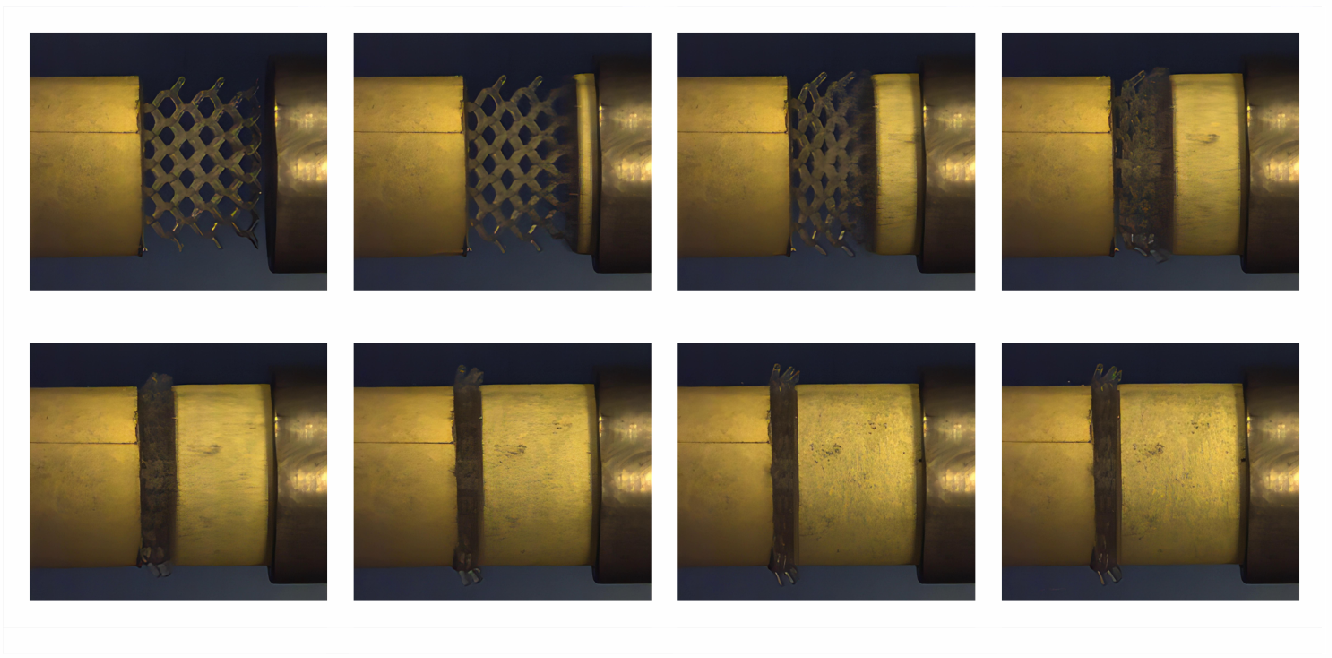}};
\draw(1.7,4.4) node[anchor=south west]{0 $\mu$s};
\draw(6,4.4) node[anchor=south west]{200 $\mu$s};
\draw(10.3,4.4) node[anchor=south west]{400 $\mu$s};
\draw(14.6,4.4) node[anchor=south west]{600 $\mu$s};
\draw(1.7,0.25) node[anchor=south west]{800 $\mu$s};
\draw(6,0.25) node[anchor=south west]{1000 $\mu$s};
\draw(10.3,0.25) node[anchor=south west]{1200 $\mu$s};
\draw(14.6,0.25) node[anchor=south west]{1400 $\mu$s};
\end{tikzpicture}}
\caption{Different states of deformation during the experiment with specimen  made of TPU like material (here \textsc{Diamond} $f_\text{V}$ = 0.2)}
\label{specimen_break_TPU_diamond}
\end{figure*}

The tests carried out induce a strong compression of the specimens. The investigated structures were compressed between 75 \% (\textsc{TruncOcta} $f_\text{V} = 0.3$) and 82 \% (\textsc{Diamond} $f_\text{V} = 0.2$). Despite this strong compression, the structures relaxe completely to their initial state within some minutes. Fig. \ref{specimen_break_TPU_octet} and Fig. \ref{specimen_break_TPU_diamond} show the compression of two specimens at different steps of time.

\subsection{Attenuation of the pulse}
The pulses are attenuated by the lattice structures and stretched over time. The amplitudes of the strain pulse also have a time delay and, in particular,  their magnitude is lower. Reason for this damping effect is the folding of the individual lattice layers. The peak of the pulse occurs when the entire structure is folded together.

\begin{figure*}[htb]
\centering
\includegraphics[width=1\textwidth]{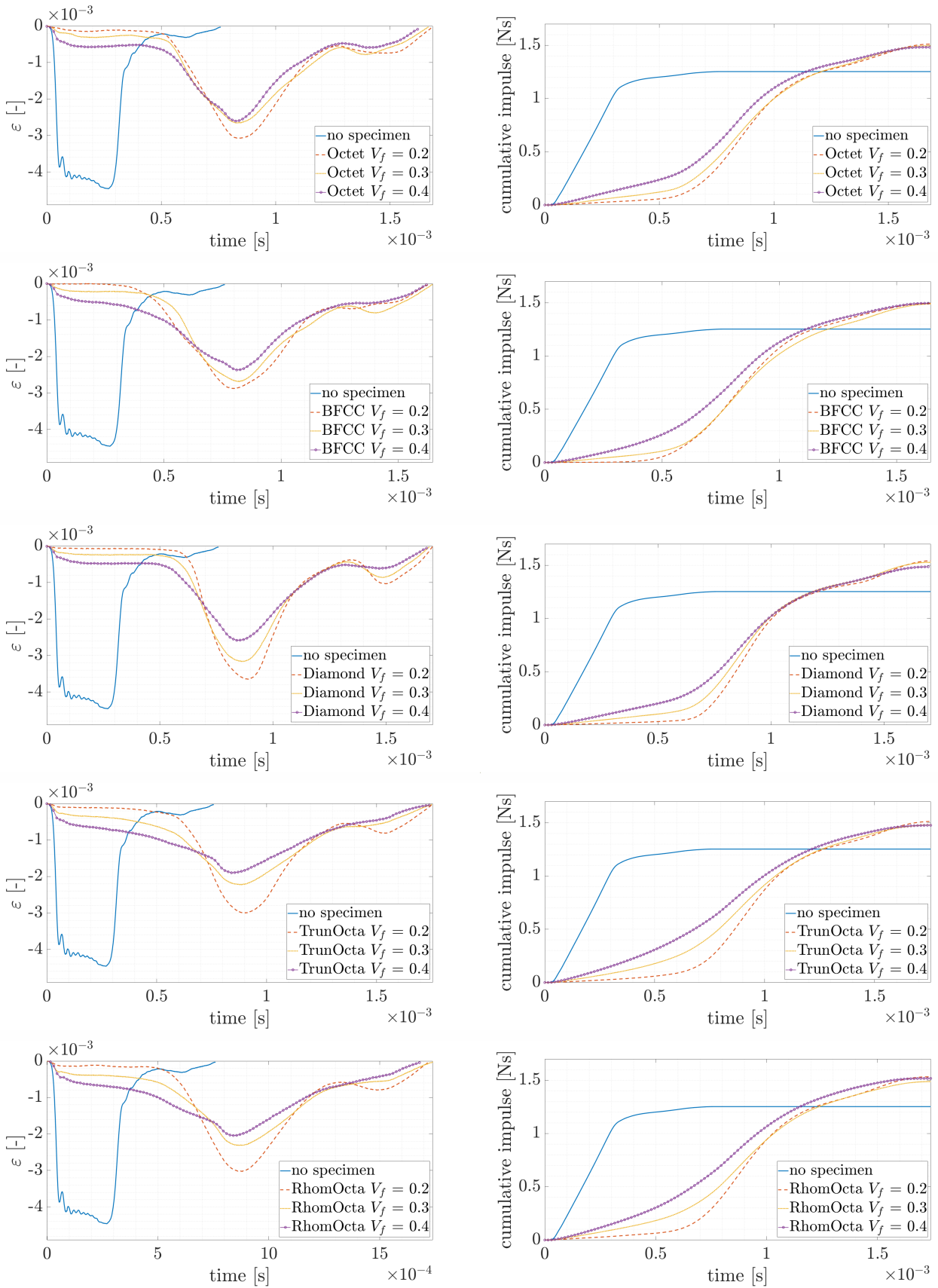};
\caption{Measured pulses signals (left) and the cumulative impulse (right) of the different lattice types}
\label{strain_cumulative_pulse}
\end{figure*}

Independent of the structure type, the amplitude of the pulse ``flattens"  as the volume fraction increases; see Fig. \ref{strain_cumulative_pulse}~(left).
For display, the recorded pulses in Fig.~\ref{strain_cumulative_pulse} are shifted to the origin.

The cumulative-impulse over time curves are obtained by integrating the impact force over time,
\begin{align}
p(t) = \int_{t_0}^{t_1} \, \varepsilon_I(t) \, E_\text{b} \frac{\pi d_\text{b}^2}{4}\, \mathrm{d}t \label{eq:cumulative_pulse}
\end{align}
where $d_\text{b}$, $E_\text{b}$ are  diameter and Young's modulus of the bar and $\varepsilon_I$ the measured signal from the strain gauges of the incident bar. The corresponding curves are shown in Fig.~\ref{strain_cumulative_pulse}~(right).

\textbf{We observe:}
that the rise in the curve starts later for specimens with the  low volume fraction of $f_\text{V} = 0.2$. However, it has a higher slope and reaches  about the same impulse. Specimens with a volume fraction of $f_\text{V} = 0.4$ do not show such an abrupt increase. The rise is slower here, representing a lower load on the material. The significantly smoother impulse transmission with an applied lattice structure into the incident bar shows a less abrupt impulse change in time than with an impact without an applied lattice structure.

The peaks of the pulses are considerably reduced by all  lattice structures. An amplitude reduction of 57\,\% is achieved by the \textsc{TrunOcta} structure with a volume fraction of $f_\text{V} = 0.4$, while the \textsc{Diamond} structure has a comparatively low reduction of 18\,\% with $f_\text{V} = 0.2$. The \textsc{Diamand} structure shows a correspondingly lower energy absorption. The structure is less stiff and deforms faster or more strongly. Correspondingly, the speed of the striker is slowed down less, which means that the higher kinetic energy of the striker is transferred to the incident bar when the \textsc{Diamond} structure is completely compressed.

When considering the pulse length, all structures stretch it over time. The stretch ranges from 113\,\% for the \textsc{BFCC} with $f_\text{V} = 0.4$ to 138\,\% for the \textsc{TrunOcta} with $f_\text{V} = 0.2$. All specimens, independent of the volume fraction $f_\text{V}$, returned to their original shape after a period of time.
We remark that no external cracks or fractures were detected in the structures.

\subsection{Comparison of frequency spectrum}
\begin{figure*}[htb]
\centering
\includegraphics[width=1\textwidth]{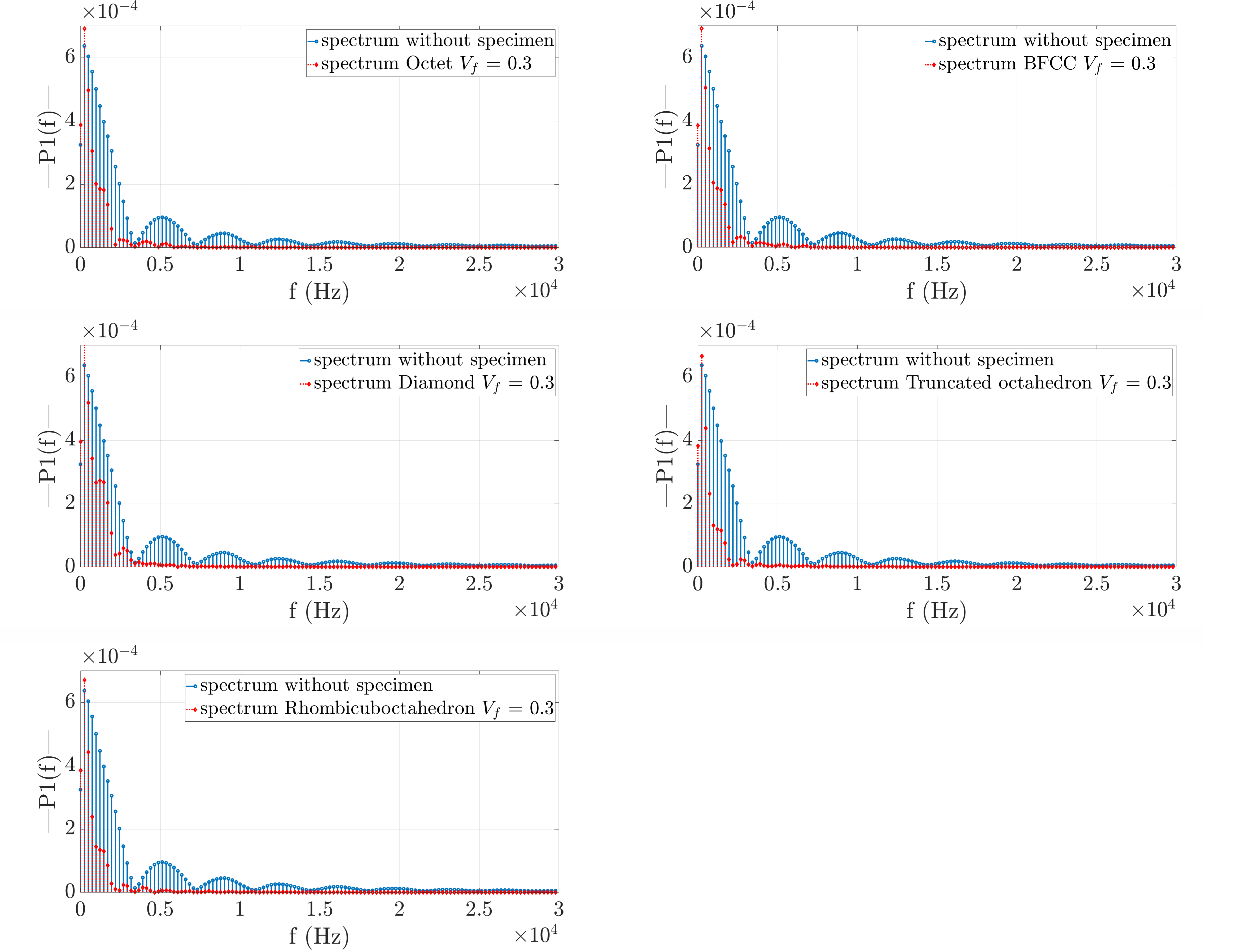};
\caption{Spectra of different lattice types with $V_f$ = 0.3}
\label{fig:spectra_02}
\end{figure*}

Observing the spectrum of the pulses provides information about the frequencies which are attenuated or amplified by the lattice structures.
The spectra of the recorded pulses are calculated with a Fast Fourier Transformation (FFT) and displayed in Fig. \ref{fig:spectra_02}.
For all lattices with $f_\text{V} = 0.3$ the recorded spectra with and without damping specimen are compared there. We observe that the lattice  structures filter out higher frequencies from the signal. The applied structures almost entirely attenuate frequencies above 7 kHz. The structures can, therefore, also be used as a low-pass filter.

\subsection{Energy absorption}
The energy dissipation through a specimen on a conventional SHB can be calculated from the difference of the work done by the incident wave $W_{\text{I}}$ to the
the sum of the reflected and transmitted pulses' work $ W_{\text{R}}, W_{\text{T}}$. This gives the work absorbed by the specimen, $\Delta W = W_{\text{I}} - \left(W_{\text{R}} + W_{\text{T}}\right)$, cf.  \cite{bieler2021investigation,beccu1987transmission,lundberg1976split}.
The work or energyis determined by integration the respective pulses,
\begin{align}
W_{\text{I}} 
= A_\text{b} \, c_\text{b} \, E_\text{b} \,  \int_{0}^t \varepsilon_{\text{I}}^2(\tilde{t}) \, \mathrm{d}\tilde{t} \,,\label{eq: energy_absorption_pulse}
\end{align}
where $A_\text{b}$ and $c_\text{b}$ are the cross-section area of the bar and its wave propagation velocity, see Table \ref{tab:SHPB_dimensions}
for the corresponding values. However, in our experiments, only the incident pulse is considered. Therefore, the
absorbed energy is determined from the difference of the reference measurement $W_{\text{ref}}$ without applied lattice structures and the calculated energy with applied lattice structures $W_{\text{lattice}}$,
\begin{align}
\Delta W = W_{\text{ref}} - W_{\text{lattice}} \label{eq:diff_W_modified_SHPB}
\end{align}
For the reference measurement, the energy is $ W_{\text{ref}}=11.37$\,Nm at a striker velocity of 15\,m/s.

\begin{figure*}[htb]
\centering
\resizebox{1\textwidth}{!}{
\begin{tikzpicture}
\draw (0,0) node[anchor=south west]{
\includegraphics[width=1\textwidth]{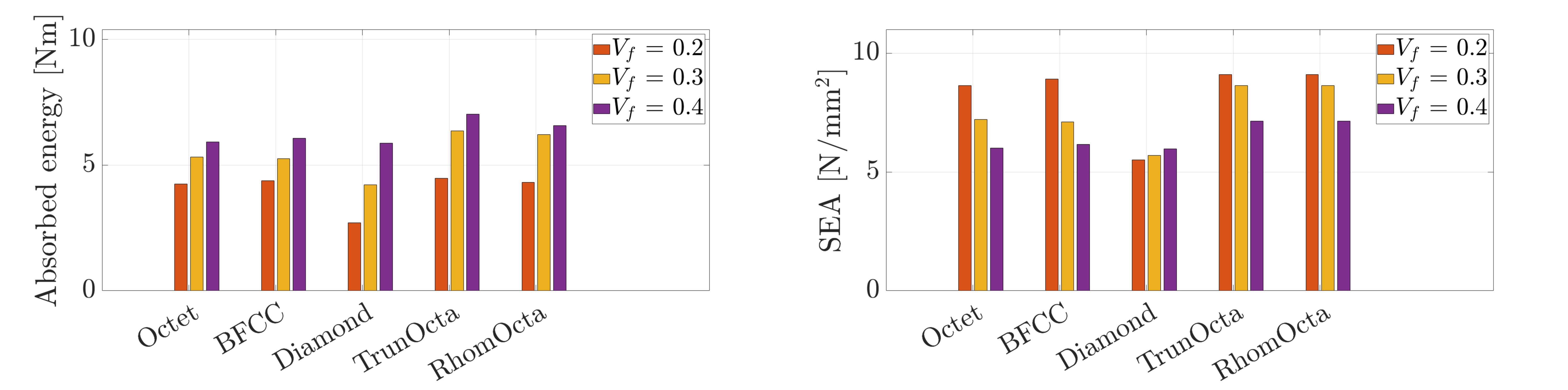}};
\draw(0.2,-0.3) node[anchor=south west]{(a)};
\draw(9,-0.3) node[anchor=south west]{(b)};
\end{tikzpicture}}
\caption{a)Total absorbed energy and b) the specific energy absoption for each lattice typ at different $V_f$}
\label{absorbed_energy_SEA}
\end{figure*}

Fig. \ref{absorbed_energy_SEA}~a) shows the total absorbed energies for the different specimens. Structures with a volume fraction of $f_\text{V} = 0.4$ have the highest total energy absorption within the same structure set. In addition, Fig.~\ref{absorbed_energy_SEA}~b) shows the specific energy absorption (SEA). This is the total energy absorption refererd the volume of the specimen,
\begin{align}
\text{SEA} = \frac{\Delta W}{V}  \label{eq:SEA}
\end{align}
which gives a more pronounced information about the absorption properties of the lattice structure.

We observe, that structures with lower volume fractions have higher specific energy absorption. The SEA decreases continuously with increasing volume fraction for all specimen types except for the \textsc{Diamond} structure. Here the SEA increases with increasing $f_\text{V}$.

The results show that the \textsc{TrunOcta} structure offers the best energy absorption among the structures presented here. However, the \textsc{RhomOcta} structure performs similarly well. Both structures are characterized by a high negative Maxwell number $M$ (\textsc{RhomOcta}: $M=-18$ and \textsc{TrunOcta}: $M=-30$). Furthermore, the structures have neither nodes in the center nor the corners of the unit cell. 
The \textsc{Octet} and the \textsc{BFCC} structures show similar properties in terms of energy absorption across all volume fractions. The \textsc{Diamond} structure also shows similar values for a volume fraction of $f_\text{V} = 0.4$, but recognizably lower energy absorption for the lower volume fractions. 
When comparing the specific energy absorption, it is immediately noticeable that there is hardly any change in energy absorption for the \textsc{Diamond} structure for the different volume fractions. All other structures show the highest SEA at the lowest volume fractions with a recognizable downward trend for higher volume fractions $f_\text{V}$.

\section{Numerical simulation of impact test} \label{sec:simulation}


In order to better understand the deformation during an impact test, numerical simulations were carried out with \textsc{Abaqus} in addition to the experiments. For this purpose, the structure of the SHPB was created without the transmission bar in \textsc{Abaqus/ Explicit}. Due to the fast deformation, problems quickly arise in the simulation about the ratio of deformation speed and wave propagation speed. Mass scaling prevented this problem by setting the target time increment to $t_{inc} = 1E^{-10}$. This makes the calculations very time-consuming and computationally intensive. For the simulation of the \textsc{Octet} structure presented here, the cluster of the University of Siegen was used on three nodes with 64 cores each and 240 GB RAM each to carry out the simulation. The simulation time was limited to 1.5 ms.

\begin{figure*}[htb]
\centering
\resizebox{1\textwidth}{!}{
\begin{tikzpicture}
\draw (0,0) node[anchor=south west]{
\includegraphics[width=1\textwidth]{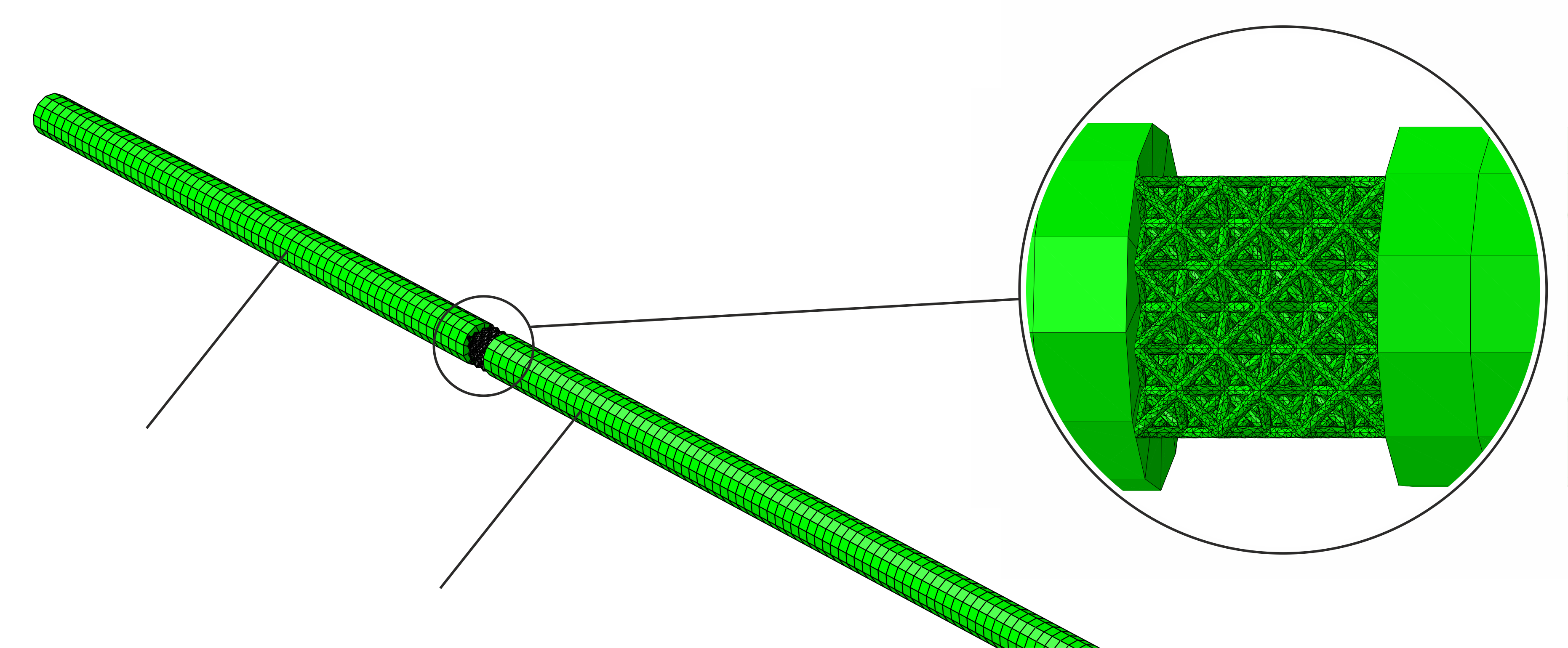}};
\draw(1.1,1.9) node[anchor=south west]{striker};
\draw(4.1,0.2) node[anchor=south west]{incident bar};
\draw(13.2,0.5) node[anchor=south west]{\textsc{Octet} specimen};
\end{tikzpicture}}
\caption{Finite element mesh for the components of the simulated components for the impact test}
\label{fig:simulation_mesh}
\end{figure*}
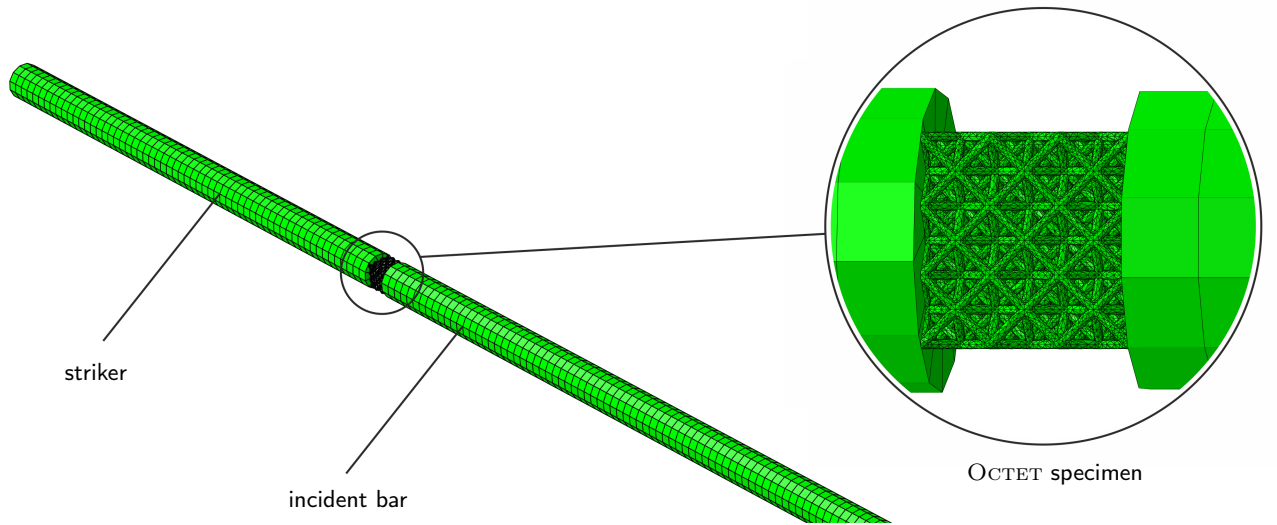

\begin{figure*}[htb]
\centering
\resizebox{1\textwidth}{!}{
\begin{tikzpicture}
\draw (0,0) node[anchor=south west]{
\includegraphics[width=1\textwidth]{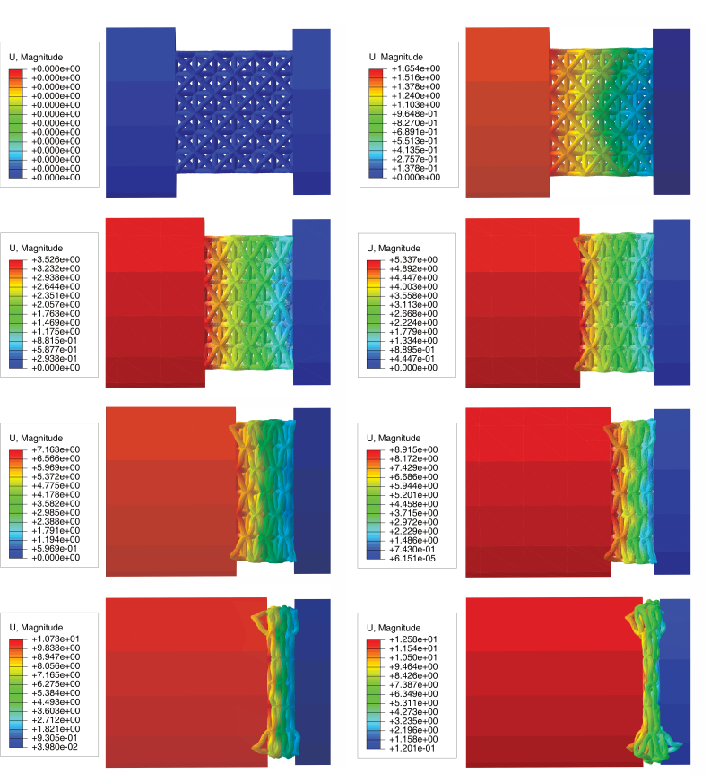}};
\end{tikzpicture}}
\caption{Numerical Simulation of the modified SHPB experiment with an \textsc{Octet} structure in \textsc{Abaqus} at different points of deformation; on the left side the striker impacts the structure}
\label{fig:simulation_def}
\end{figure*}

The incident bar and the striker were meshed with hexagonal elements and only allowed movements in the axial direction to simulate the support of the bars in the experiment. Due to the complexity of the lattice structures, these were meshed with ten-node modified quadratic tetrahedron elements (C3D10M); see Fig. \ref{fig:simulation_mesh}. Surface-to-surface contact with a friction coefficient of 0.15 was chosen for the contact between the incident bar and the specimen and between the striker and the sample. In addition, self-contact was chosen for the lattice structure to map large deformations within the lattice. The striker velocity was analogous to the experiments via Predefined field velocity of $v_{\text{st}} = 15$ m/s. The simulation shows realistic deformation and good agreement with the experiments; see Fig. \ref{fig:simulation_def}.

\section{Discussion} \label{sec:discussion_summary}
The current study presents the energy absorption of different additively manufactured lattice structures under impact loads. The loads were realized by a modified SHPB. For each structure, a unit cell was developed for a predefined length. By varying the strut thickness, the volume fraction $f_{\text{V}}$ could be adjusted for each structure. The different structure types were each created for three volume fractions and then compared with each other in terms of their energy absorption properties.

The results obtained in this study show that lattice structures can actively contribute to the protection of applications or people. Particular attention was paid to the aspect of the sustainability of such structures, i.e. multiple loads. By choosing a suitable material, the energies caused by an impact could be greatly reduced. The structures show a complete return to their original state after a short recovery time. This is made possible by folding the individual layers, which in turn causes the impulse to be stretched. This flattens the pulse rise, which in turn reduces the pulse's amplitude.

It was found that the \textsc{RhomOcta} and the \textsc{TrunOcta} structures show the best energy absorption of the investigated structures. Both structures have a negative Maxwell number $M$ and have no nodes in the corners of the unit cells. However, the \textsc{Diamond} structure also turns out to be remarkable, as it shows equally effective properties for all investigated volume fractions $f_{\text{V}}$ when considering the specific energy absorption SEA.

In addition, the pulses were analyzed using FFT to determine their frequencies. This revealed a substantial attenuation of higher frequencies when using the lattice structures. As a result, applied structures also act as a low-pass filter.

For the volume densities tested in this investigation, it was found that the energy absorption for the lattice structure increases with increasing volume density. For the \textsc{Diamond} structure, the specific energy absorption SEA remains constant for the investigated volume fractions $f_{\text{V}}$. Future investigations with further volume fractions $f_{\text{V}}$ of the  \textsc{Diamond} structure should be carried out to substantiate this finding.

A numerical simulation was carried out with \textsc{Abaqus}, which shows good agreement with the deformation from the experiments. Due to the high computational time, the regeneration of the structure after loading could not be simulated within this work. At this point, further investigations or simulations are necessary in order to achieve further understanding in this regard.

\section{Appendix} \label{sec:appendix}

\begin{table*}[htb]
\caption{Required strut radius $r$ for a given volume fraction $V_f$ as well as the resulting surface area $A$ of the unit cells ($l=4.5$ mm)}
\label{tab:volume_fraction} 
\begin{tabularx}{\textwidth}{lXXX} \toprule
\noalign{\smallskip}
 & $f_\text{V}$ & $r$ [mm] & $A$ [mm$^2$]\\
\noalign{\smallskip}
\hline
\noalign{\smallskip}
\textsc{Octet} & 0.2 & 0.308 & 145.98\\
\noalign{\smallskip}
\textsc{BFCC} & 0.2 & 0.359 & 122.49\\
\noalign{\smallskip}
\textsc{TrunOcta} & 0.2 & 0.442 & 100.46\\
\noalign{\smallskip}
\textsc{RhomOcta} & 0.2 & 0.335 & 124.28\\
\noalign{\smallskip}
\hline
\noalign{\smallskip}
\textsc{Octet} & 0.3 & 0.390 & 168.65\\
\noalign{\smallskip}
\textsc{BFCC} & 0.3 & 0.456 & 141.36\\
\noalign{\smallskip}
\textsc{TrunOcta} & 0.3 & 0.564 & 116.43\\
\noalign{\smallskip}
\textsc{RhomOcta} & 0.3 & 0.428 & 142.45\\
\noalign{\smallskip}
\hline
\noalign{\smallskip}
\textsc{Octet} & 0.4 & 0.466 & 183.37\\
\noalign{\smallskip}
\textsc{BFCC} & 0.4 & 0.546 & 153.41\\
\noalign{\smallskip}
\textsc{TrunOcta} & 0.4 & 0.679 & 127.05\\
\noalign{\smallskip}
\textsc{RhomOcta} & 0.4 & 0.518 & 153.08\\
\noalign{\smallskip}
\hline
\end{tabularx}
\end{table*}

\begin{table*}[htb]
\caption{Dimensions of the SHPB}
\label{tab:SHPB_dimensions} 
\begin{tabularx}{\textwidth}{lXX} \toprule
\noalign{\smallskip}
 & incident bar \& transmission bar & striker\\
\noalign{\smallskip}
\hline
\noalign{\smallskip}
length $l$ [mm] & 3000 & 300\\
\noalign{\smallskip}
diameter $d$ [mm] & 20 & 20\\
\noalign{\smallskip}
density $\rho$ [kg/m$^3$] & 1180 & 1180\\
\noalign{\smallskip}
Young's modulus $E$ [MPa] & 3000 & 3000\\
\noalign{\smallskip}
wave speed $c$ [m/s]& 2500 & 2500\\
\noalign{\smallskip}
\hline
\end{tabularx}
\end{table*}


\printcredits

\section*{Declaration of competing interest}
The authors declare that they have no competing financial interests or personal relationships that could have appeared to influence the work reported in this paper.
\section*{Data availability}
Data will be made available on reasonable request.
%




\end{document}